\documentstyle[prb,aps,preprint]{revtex}
\begin{document}
\draft
\title{Bosonization on the lattice: the emergence of the higher
harmonics}
\author{J. Ferrer\cite{email1}}
\address{Departamento de F\'{\i}sica de la Materia Condensada C-12,
Universidad Aut\'onoma de Madrid, 28049 Madrid, Spain}
\author{J. Gonz\'alez\cite{email2}}
\address{Instituto de Estructura de la Materia, Serrano 123,
28006 Madrid, Spain}
\author{and M.-A. Mart\'{\i}n-Delgado\cite{email3}}
\address{Departamento de F\'{\i}sica Te\'orica I, Universidad
Complutense de Madrid, 28040 Madrid, Spain}

\maketitle

\begin{abstract}
A general and transparent procedure to bosonize fermions placed on
a lattice is presented. Harmonics higher than $k_F$
in the one-particle Green function
are shown to appear due
to the compact character of real electron bands. Quantitative
estimations of the role of higher harmonics
are made possible by
this bosonization technique.
\end{abstract}


Bosonization methods have provided us with a thorough understanding
of the  physics of
$1+1$  dimensional model systems in several branches of Theoretical
Physics  as, for example,
Condensed Matter \cite{solyom,shankar}.
The earliest implementation of such techniques dates back to the sixties,
when Luttinger \cite{lutt} proposed his model, which was subsequently
solved partially
by Mattis and Lieb \cite{mattis1}. Ten
years later, Luther and Peschel, and Mattis \cite{luther,mattis2}
made the picture more concrete by combining their refined
version of bosonization with known results for some models previously
solved by Bethe
ansatz. Soon afterwards, and
independently, a similar boson-fermion equivalence was also obtained
by Coleman \cite{coleman}
in the
context of the Sine-Gordon model. Perhaps the most pictorial representation
of a fermion in terms of bosons has been given by
Mandelstam \cite{mandelstam}.
According to it,  fermions should be understood in a purely
bosonic theory as soliton operators interpolating between
different particle vacua. Reciprocally, the generic behavior of
electron liquids in $1+1$ dimensions
is such  that all the excitations of the Fermi sea
can be classified into a set of boson operators.

The paradigm of a theory which can be solved by means of
bosonization  is the Luttinger model. This is a one dimensional
model, in which electrons interact only through
density operators of definite chirality. The
total Hamiltonian can be expressed as a quadratic form of two
boson fields with opposite chiralities, and this fact renders the model
completely  integrable. There are, however, some assumptions in
the Luttinger model which make of it an approximate description  of the
physics of real electrons. The most important of them  are made by
considering a perfect linear dispersion relation for
the original electrons, and by supposing
that the two branches (corresponding
to the two different Fermi points) can be artificially extended
{\em ad infinitum} in both directions (see
fig. 1(a)). Obviously, the infinite
collection of states deep inside the Fermi sea is not
present at all in a real physical situation, and
the hypothesis that they do not modify the essential
properties learned from  Luttinger's model becomes crucial.

Let us explain with more detail some of the complications which appear
in real condensed matter systems and which blur the sharp picture
brought about by the bosonization of the Luttinger model.
First, the dispersion relation for one species of free
non-relativistic  fermions
is simply a parabola, $\epsilon(p)=p^2/2m$, which is bounded from below,
but not from above (fig. 1(b)).
Second, when these formerly free electrons are placed in a lattice
and interact with the periodic substrate potential of the atoms, the
dispersion relation  becomes  also bounded from above, and peaks, in the
simplest case, at the Bragg point so that we are left with a
compact band of width $D$ (see fig. 1(c)).
The aforementioned phenomena bear two related effects.
One is the appearance, due to the compact character of the band, of
chirality breaking processes, which mix both branches.
This is tantamount to say that one can not divide the physical
electron field operator into
right and left moving pieces unambiguously.
The other effect is the curvature of the band.
We argue that in most cases the
first effect  is the most relevant one because it is
the source of the appearance of higher harmonics
in electron correlation functions, while the second one gives
rise to harmless renormalizations of the parameters.
In the case of the electron Green function, for instance,
together with the naive frequencies $\pm k_F$ one expects
higher modulations at $\pm 3 k_F$, $\pm 5 k_F$, etc :
\begin{equation}
\langle \Psi (x) \Psi^{+} (y) \rangle =
\sum_{n = 0}^{\infty } c_n \frac{i}{(x - y)^{\alpha_n }}
    e^{i (2n + 1) k_F (x - y)}   \;\;\; + \;\;\; h.c.
\end{equation}

An underlying assumption of the bosonization technique is that only
long wavelength
fluctuations of the density of particles affect the physics of the
problem.  One is therefore
allowed to average all magnitudes over distances much larger than the
average distance among
particles, $r_s \simeq 1/k_F$ ---in particular, the commutation
relations  and expectation
values of density operators.
We proceed now to describe qualitatively the effects of a finite
density of electrons on the
accuracy of the mapping of electrons to bosons.
\begin{itemize}
\item
For intermediate densities, when $k_F$ is placed approximately
in the middle of the band,
its curvature is  very small and we can linearize it around the Fermi points
with a high degree of accuracy.  On the other hand, chirality-breaking
processes have high
energy and can be disregarded as a first approximation.
\item
At low densities, $k_F$ is close to the bottom of the band.
In this  case, the discrete
character of the particles is important, because the length scale set
by $r_s$ is large. One should add that the processes
which break the chirality have low energies and need to be taken
into account. They show up in the form of higher harmonics
in the electron correlation functions.
\item
At high densities the Fermi wave-vector is close to the top of the band,
and due to particle-hole symmetry the qualitative discussion
in the former item applies.

\end{itemize}

The issue of the emergence of higher
harmonics is not new, and has been considered before
mainly in connection with the effects of the curvature of the
dispersion relation. A quite original point of view,
due to Haldane\cite{haldane}, rephrases the problem as the
incorporation of the discrete nature of the particles into  the
bosonization program. In either case, it becomes clear that the
boson-fermion transcription valid within the Luttinger model should
be corrected to take into account more realistic dispersion
relations. In particular, if we consider a {\em compact}
dispersion relation and give up the perfect division between
left and right movers of the Luttinger model, hybridization
effects between the two chiral fields will appear giving rise to
higher harmonic modulations in the electron correlation
functions. Up to date, though, there has been no attempt to
understand what is technically the source of such hybridization
and, thereafter, to propose a systematic way of correcting the
original boson expression of the fermion operator.

The purpose of the present paper is to incorporate the compact
character of the band ---i.e.
the chirality-breaking processes--- into the bosonization technique
in a non-perturbative,
and essentially exact, way. Our goal is to set up a scheme
that permits to study {\em
quantitatively} the role of higher harmonics. These might
be relevant for the physical
behavior of some experimental devices, e.g.: quantum wires
\cite{fisher,schulz}.
It is our belief that our procedure places bosonization in the doorway of
quantitative computations of response functions of one-dimensional
systems in Condensed
Matter.

We begin with a reminder of the main lines of the simple
bosonization program,
where the assumption of an infinite linear dispersion relation
of the two electron branches becomes essential. In the
Luttinger model the electronic spectrum of the ``free''
Hamiltonian is that represented in fig. 1(a). There are two types
of fermion modes, say $a_{k}, a_{k}^{+}$ and
$b_{k}, b_{k}^{+}$, for the respective right and left branches
of the spectrum. It is well-known that the only excitations
supported by the Fermi sea of fig. 1(a) are density fluctuations
of the form
\begin{equation}
\rho_{kR} = \sum_{q} a^{+}_{q+k} a_{q}
\end{equation}
for the right branch, and
\begin{equation}
\rho_{kL} = \sum_{q} b^{+}_{q+k} b_{q}
\end{equation}
for the left branch.
There are obviously other fluctuation processes in which
electrons are transferred from one branch to the other, but in
the Luttinger model they amount to the introduction of a
conserved quantum number $J$. The important point is that the
above currents satisfy the commutation relations
\begin{eqnarray}
\left[ \rho_{- \tilde{k} R} , \rho_{k R} \right] & = & \delta_{k
\tilde{k} } \: k \: \frac{L}{2 \pi }     \nonumber \\
\left[ \rho_{- \tilde{k} L} , \rho_{k L} \right] & = & - \delta_{k
\tilde{k} } \: k \: \frac{L}{2 \pi }   \label{l}
\end{eqnarray}
where $L$ is the length of the dimension in which the electrons
are confined. The linear dependence of the commutators
(\ref{l}) can be
rigorously proved under the hypothesis of an infinite linear
dispersion relation as shown in fig. 1(a). It allows us to define
boson creation and annihilation operators
\begin{eqnarray}
B^{+}_{k} = \sqrt{ \frac{2 \pi}{L \left| k \right| }}\, \rho_{k R} \;\;\;
  k > 0  &  ,  &\:
B^{+}_{k} = -\sqrt{ \frac{2 \pi}{L \left| k \right| }} \, \rho_{k L} \;\;\;
  k < 0    \nonumber \\
B_{k} = \sqrt{ \frac{2 \pi}{L \left| k \right| }} \, \rho_{- k R} \;\;\;
  k > 0   &  ,  &\:
B_{k} = -\sqrt{ \frac{2 \pi}{L \left| k \right| }} \, \rho_{- k L} \;\;\;
  k < 0  \nonumber\\&&
\label{bop}
\end{eqnarray}
which satisfy perfect canonical  commutation relations
\begin{equation}
\left[ B_k , B^{+}_{\tilde{k}} \right] = \delta_{k \tilde{k} }
\label{ccrl}
\end{equation}
These oscillators can in turn be assembled into two chiral
boson fields
\begin{eqnarray}
\Phi_R (x) & = & \frac{2 \pi }{L}  \left(
    \, x \, N_R + i\, \sum_{k \neq 0 }
    \,   \frac{   e^{- i k x }  }{k}
 \rho_{k R}    \right)      \label{rb} \nonumber\\
\Phi_L (x) & = & \frac{2 \pi }{L}  \left(
    \, x \,N_L + i\, \sum_{k \neq 0 }
   \,   \frac{   e^{- i k x }  }{k}
  \rho_{k L}    \right)        \label{lb}
\end{eqnarray}
Here $N_R$ and $N_L$ are the normal ordered charges
for the respective channels. This boson codification of the
electron excitations is only half of the boson-fermion
equivalence. It can  also be shown that the fermion field may
be expressed in terms of the above boson fields. In particular,
a correct representation for the
two fermion chiralities is
\begin{eqnarray}
\Psi_R (x)  & = & : e^{i \Phi_R (x) } :  \nonumber\\
\Psi_L (x)  & = & : e^{-i \Phi_L (x) } : \label{rpl}
\end{eqnarray}
These are the expressions for the soliton
(fermion) annihilation operators found by Mandelstam.
They have the virtue of
satisfying the equal-time canonical anticommutation relations of
fermion operators. Finally and more important, the representation
(\ref{rpl}) reproduces the form of the fermion
correlators
\begin{eqnarray}
\langle \Psi_R (x) \Psi^{+}_R (x') \rangle  & = &
\frac{i}{x - x'}   \label{rcorr}    \\
\langle \Psi_L (x) \Psi^{+}_L (x') \rangle  & = &
\frac{-i}{x - x'}
\end{eqnarray}

At this point we undertake the analysis of how this program has
to be modified when a more realistic, compact dispersion relation
is considered in the description  of the electronic system.
Electrons usually feel the background periodic potential of the
atomic lattice. This substrate potential changes their parabolic
dispersion relation into a band (figs. 1(b) and 1(c)).
Despite the fact that in such case no natural distinction between
right and left modes can be made, we want to keep the
separation into two different branches for computational
purposes. In fact, even in the case of a compact spectrum of the
kind shown in  fig. 1(c) the static electron correlator still
shows two different modulations corresponding to the left and
right branches. Suppose, for instance, that
we write the mode expansion for
the fermion field
\begin{equation}
\Psi (x) = \frac{2 \pi}{L} \sum_{k = - \pi}^{0} \,e^{i kx} b_{k} +
    \frac{2  \pi}{L}  \sum_{k = 0}^{\pi}  \,e^{i kx}  a_{k}
\end{equation}
Then, a straightforward computation gives the result (in the
limit $L \rightarrow \infty $)
\begin{eqnarray}
\langle \Psi (x) \Psi^{+} (y) \rangle & = &
   \int_{- \pi}^{- k_F } dk \: e^{i k(x - y) } +
    \int_{k_F }^{\pi } dk \: e^{i k(x - y) } \nonumber   \\
  & =  &   \frac{i}{x - y} \,e^{i k_F (x - y)}  +
       \frac{-i}{x - y} \,e^{-i k_F (x - y)}   \label{corr}
\end{eqnarray}
This is exactly the same expression that one obtains for the
static correlator in the Luttinger model. However, as we are
going to see, the bounded character of the spectrum of boson excitations
requires appropriate modifications in the intermediate steps
which lead to (\ref{corr}) within the bosonization approach.

It is worthwhile to remark that the particular energy values of
the electron modes are irrelevant for the purpose of computing
the static correlators. The only important point is that the
Fermi sea comprises now a connected set of states from $k
= - k_F$ to $k = k_F$. Given that we do not have an infinite
dispersion relation anymore, we would like to write tentatively
the set of chiral currents
\begin{eqnarray}
\rho_{kR} & = & \sum_{0 < q+k, q < \pi} a^{+}_{q+k} \,a_{q} \nonumber \\
 \rho_{kL} & = &  \sum_{- \pi < q+k, q < 0} b^{+}_{q+k} \,b_{q}
\end{eqnarray}
The first exercise in order to test the bosonization
procedure is to check the linear dependence of the
commutator of currents with like chirality:
\begin{eqnarray}
 \left[  \rho_{-kR} , \rho_{kR}  \right]  & = &
  \left[ \sum_{0 < q-k, q < \pi} a^{+}_{q-k}\, a_{q}  ,
    \sum_{0 < r+k, r < \pi} a^{+}_{r+k}\, a_{r} \right]\nonumber   \\
&&\nonumber\\
  & = & \sum_{0 <q-k, r<\pi }
      \delta_{q,r+k} a^{+}_{q-k} a_{r}-\nonumber\\
     &&\:\:\:\:\:\:
     - \sum_{0<r+k,q<\pi }
       \delta_{r,q-k} a^{+}_{r+k} a_{q}  \label{sum}
\end{eqnarray}
It is important to realize that in these sums all the
subindices run from $0$ to $\pi$. For this reason, one can see
that for sufficiently small values of $k$ the first sum in
(\ref{sum}) has $L/(2 \pi)$ times $k$ more contributions than
the other. This agrees with the linear dependence in (\ref{l}).
However, in the case of a band less than half-filled,
when $k > k_F$ there are not enough excitations of the
Fermi sea and the commutator remains equal to
$L/(2 \pi) \,k_F$, up to a
value of $k = \pi - k_F$. From that value it begins to decrease
linearly and reaches $0$ at $k = \pi$. This picture is valid, as we
have said, for values of $k_F$ between $0$ and $\pi/2$. When the
band is more than half-filled we get a similar form of the
commutator but with linear growing up to $\pi - k_F$ and later
linear decrease from $k_F$ to $\pi$.

We may pause at this point and think of the physical reasons for
this deviation of the commutator from a perfect linear dependence.
They can indeed be found by looking
at the very essence of the computation
performed above. As a matter of fact,
{\em the value
of the commutator is a measure of the number of available
one-particle excitations over the Fermi sea}. In the case of a band less
than half-filled, for instance, it is clear that for small
values of momentum transfer $k$ there is no problem in exciting
$L/(2 \pi)\, k$ electrons from below the Fermi level to
states above it. When $k > k_F$, though, we cannot continue
pulling out right modes once we reach the bottom of the band and the
number of available excitations is less than $L/(2 \pi) \,k$.
This argument explains also why the actual number remains
constant and equal to $L/(2 \pi) \,k_F$, up to a momentum transfer
$\pi - k_F$.

However, this clear interpretation of the
functional dependence of the commutator also shows that the
present picture is physically incorrect. In fact, it is
only our artificial division between right and left modes what
has prevented us from considering another set of admissible
one-particle excitations for $k > k_F$. These correspond to the
transfer of electrons below the Fermi level in the range $[-k_F
, 0]$ to states above the Fermi level in the right branch.
Obviously, there is no reason for not considering these
excitations on the same footing than those taken into account
before within the same branch. Bearing this in mind, it seems
more natural the definition of  the currents
\begin{eqnarray}
\rho_{kR} & = & \sum_{0 < q+k, q < \pi} a^{+}_{q+k}\, a_{q}
   + \sum_{-k_F < q < 0} a^{+}_{q+k} \,b_{q} \label{mr} \nonumber \\
 \rho_{kL} & = &  \sum_{- \pi < q+k, q < 0} b^{+}_{q+k}\, b_{q}
   + \sum_{0 < q < k_F} b^{+}_{q+k}\, a_{q}  \label{ml}
\end{eqnarray}
It is clear that the correct counting of
excitations leads to a situation in which for momentum transfer
$k = 2 k_F$ the number of them equals the maximum value
$L/(2 \pi) \,2 k_F$, including the extreme process in which an
electron slightly below the Fermi level is transferred above it at
the other Fermi point.
This value $L/(2 \pi) 2 k_F$ is also the
cutoff for the commutator. The correct
physical picture says, then, that the commutator should be a linear
function growing up to $L/(2 \pi)\, 2 k_F$ at $k = 2k_F$,
remaining constant until $k = \pi - 2k_F$ and then linearly
decreasing to $0$ at $k = \pi$.

The most important effect of the lattice is, therefore, to replace
the  commutators in (\ref{l}) by a {\em bounded}
function in the interval $[0, \pi]$,
which we will call $L/(2 \pi )\, f(k)$, i.e.:
\begin{equation}
\left[ \rho_{- k R} , \rho_{k R} \right]  =
\frac{L}{2 \pi }\:  f(k)
\end{equation}
$f(k)$ is depicted in figure 2.
This in turn modifies the properties of the bosons that one
can build from the currents  (\ref{ml}).
The correct definition of boson creation and annihilation
operators should be now
\begin{eqnarray}
B^{+}_{k} = \sqrt{ \frac{2 \pi}{L f \left(|k|\right) }}\, \rho_{k R} \;\;
  k > 0  &  ,  &\,
B^{+}_{k} = -\sqrt{ \frac{2 \pi}{L f \left(|k|\right) }}\, \rho_{k L} \;\;
  k < 0    \nonumber \\\nonumber\\
B_{k} = \sqrt{ \frac{2 \pi}{L f \left(|k|\right) }}\, \rho_{- k R} \;\;
  k > 0   &  ,  &\,
B_{k} = -\sqrt{ \frac{2 \pi}{L f \left(|k|\right) }} \,\rho_{- k L} \;\;
  k < 0   \nonumber\\&&
\end{eqnarray}
in order to preserve the canonical commutation relations (\ref{ccrl}).
We want to maintain at this point the relation that exists in
the Luttinger model
between the fields $\rho_L (x) , \rho_R (x)$
and $\Phi_L (x) , \Phi_R (x)$
\begin{equation}
\rho_L (x) = \frac{L}{2 \pi} \,\nabla \Phi_L (x)  , \;\;\;\;\;\;\;
  \rho_R (x) = \frac{L}{2 \pi} \,\nabla \Phi_R (x)
\label{rel}
\end{equation}
with the only difference that now $\nabla$ is the lattice derivative.
The two boson fields
\begin{eqnarray}
\Phi_L (x) & = & i \sum_{k < 0} \sqrt{  \frac{2 \pi}{L}  }
\frac{ \sqrt{f( \left| k \right| )} }{2
\sin(k/2)} ( e^{-i kx} B_{k}^{+} -  e^{i kx} B_{k} ) \nonumber    \\
\Phi_R (x) & = & i \sum_{k > 0}  \sqrt{  \frac{2 \pi}{L}  }
 \frac{ \sqrt{f( \left| k \right| )} }{2
\sin(k/2)} ( e^{-i kx} B_{k}^{+} -  e^{i kx} B_{k} )
\end{eqnarray}
are chiral, in the sense that $\Phi_R (x)$, for instance,
creates excitations in  the forward direction and destroys them in
the backward direction. Their properties, though, are
non-standard, since they account in their structure for the
finiteness of the number of modes of the lattice. Over very
large distances, we should expect from them the same behavior
found in the Luttinger model. This is guaranteed by the fact
that the function $f(k)$ is linear in $k$ for small values  of
the argument. Over smaller distances, though, we  start to
feel the effects of the discreteness of the number of particles.

The above  considerations are exemplified  by the  computation
of the correlator
\begin{equation}
\langle  e^{ i \Phi_R (x) }   e^{- i \Phi_R (0) }  \rangle
\label{ev}
\end{equation}
which in the Luttinger model equals the fermion propagator
(\ref{rcorr}). A straightforward calculation leads,
in the limit $L \rightarrow \infty$, to
\begin{eqnarray}
\langle  e^{ i \Phi_R (x) }   e^{- i \Phi_R (0) }  \rangle &=&
exp  \left\{ - \int_{0}^{\pi} dk  \frac{f(k)}{4  \sin^2 (k/2) }
\left( 1 - e^{i kx}  \right) \right\}  \label{int}\nonumber\\
&=& e^{-I}
\end{eqnarray}
We are mainly interested in the behavior of the correlator at
large values of $x$. The evaluation of $I$
in this regime still appears to be unfeasible, but
in the limit of small $k_F$ (compared to $\pi $) we may consider
the effects of the integration over large values of $k$ as
irrelevant. We can then approximate $I$ by
\begin{eqnarray}
I& \approx &  \int_{0}^{\Lambda} dk \frac{f(k)}{k^2}
   \left( 1 - e^{i kx}  \right) \nonumber\\& = &
  \int_{0}^{2 \,k_F} dk \frac{1}{k}  \left( 1 - e^{i kx}  \right)
 + 2 \,k_F \int_{2 k_F}^{\Lambda} dk \frac{1}{k^2}  \left( 1 - e^{i kx}
         \right) \nonumber         \\
 & \approx & log( 2\, k_F x) + \gamma_E + 1 - i\,\frac{\pi}{2}
 - \frac{e^{i 2 k_F x} }{(2 k_F x)^2} + F(x, \Lambda) + \ldots
\end{eqnarray}
$\Lambda$ plays here the role of an upper cutoff, and the function
$F(x, \Lambda)$ is
\begin{equation}
F(x, \Lambda ) = -2 \,\,\frac{k_F}{\Lambda } +  2\,i\,\,
\frac{k_F}{\Lambda } \,\frac{e^{i \Lambda x}}{ \Lambda x}
\end{equation}
At small values of $k_F /\Lambda $ the influence  of the
cutoff can be disregarded and we get the
asymptotic expansion  for the correlator
\begin{equation}
\langle  e^{ i \Phi_R (x) }   e^{- i \Phi_R (0) }  \rangle =
C \,\,\frac{i}{2 k_F x} + C \,\,e^{ i 2 k_F x}\, \frac{i}{(2 k_F x)^3} +
  \ldots
\label{for}
\end{equation}
The first term corresponds to the right-handed piece of
the electron propagator (\ref{corr}), while the rest are
contributions which arise from the structure of the boson field
operators over distances corresponding to the mean separation among
particles.

The main conclusion which follows from the evaluation of
(\ref{ev}) is that the boson representation
(\ref{rpl}) of the two fermion chiralities cannot be
correct, since it produces spurious contributions to the
electron propagator as shown in (\ref{for}).
We stress again the
fact that the electron propagator is given in any event by the
expression (\ref{corr}).
It suggests that the left-right chiral
decomposition is still at work in the free theory of fig. 1(c),
showing  no
other harmonics than those at $k_F$ and $- k_F$. The structure
of the higher order contributions in (\ref{for}), in particular
the modulation at $2 k_F$, shows that the boson representation
(\ref{rpl}) can be conveniently corrected in order to
cancel out spurious terms in the fermion propagator. Actually,
it is not a coincidence that the subdominant order in
(\ref{for}) is just {\em the opposite} of the dominant contribution from
\begin{equation}
\langle e^{i \left( \Phi_L (x) + \Phi_R (x) \right)}
    e^{i \Phi_R (x)} e^{-i \left( \Phi_L (0) + \Phi_R (0) \right)}
    e^{- i \Phi_R (0)}  \rangle
\end{equation}
Thus, the correct boson representation of the chiral fermion
operators is
\begin{eqnarray}
\Psi_R (x)  & = &  e^{i \Phi_R (x) } + c_1 e^{i 2 k_F x}
    e^{i \left( \Phi_L (x) + \Phi_R (x) \right)} \,
    e^{i \Phi_R (x)}   + \ldots   \label{ccr}     \\
\Psi_L (x)  & = &  e^{-i \Phi_L (x) } + c_1 e^{-i 2 k_F x}
    e^{- i \left( \Phi_R (x) + \Phi_L (x) \right)}
    e^{-i \Phi_L (x)}   + \ldots     \label{ccl}
\end{eqnarray}
By keeping  the decomposition
\begin{equation}
\Psi (x) = e^{- ik_F x} \Psi_L (x) + e^{ ik_F x} \Psi_R (x)
\end{equation}
it is not difficult to see that the use of (\ref{ccr}) and (\ref{ccl})
reproduces the correct expression of the electron
propagator (\ref{corr}), provided that $c_1 = C^{-1}$. The form
of the corrections in (\ref{ccr}) and (\ref{ccl}) coincides with
what has been advocated by other authors \cite{haldane}. Here, we have
accomplished a quantitative derivation of them, precise enough
to determine the coefficients of the series within a given
model. Our argumentation also clarifies conceptually that it is
the bosonization method what introduces the higher harmonic
contributions, as in (\ref{ccr}) and (\ref{ccl}), though in some
instances ---like that of the free electron system---
the only modulations in
the electron propagator are at $k_F$ and $-k_F$.

We follow at this point the standard bosonization procedure, by
which the boson representation of the fermion operators remains
unchanged after switching the interaction. Thus, we are in the
position to make explicit statements regarding the structure of
the electron propagator in the interacting theory. In general
there are couplings in the interacting  hamiltonian
which mix explicitly the two chiral parts of the electron field.
This mixing has to be considered alongside with the underlying
chiral mixing already present
in the boson representation.
The interplay between them gives rise, in addition to the
standard $k_F$ modulation, to $3 k_F$ and higher order
modulation terms in the electron propagator,
as we are going to see in what
follows. It is worth to mention that the signal of the
$3k_F$ modulation has been observed
numerically by Ogata and Shiba \cite{ogata}
in the strong coupling limit of the one-dimensional Hubbard
model.

Let us take, for the sake of simplicity,
a simple $g$-ology model consisting of
forward scattering terms of $g_2$ and $g_4$ type \cite{solyom}
\begin{eqnarray}
H & = & \int dx \; i v_F ( \Psi _R ^{\dagger } \partial _x \Psi _R -
\Psi _L ^{\dagger } \partial _x \Psi _L )  \nonumber     \\
 &   &  + \;  \int dx \left\{
   g_2 \Psi _R ^{\dagger } \Psi _R  \Psi _L ^{\dagger } \Psi _L +
\frac{g_4}{2} \left[ ( \Psi _R ^{\dagger } \Psi _R )^2
     + ( \Psi _L ^{\dagger } \Psi _L )^2 \right] \right\}
\end{eqnarray}
As is well-known, in the boson representation this hamiltonian
is diagonalized by the canonical transformation
\begin{equation}
\left(  \begin{array}{c}
         \Phi_{L}  \\   \Phi_{R}
       \end{array}   \right)  =
 \left(  \begin{array}{cc}
         cosh \: \lambda  & - sinh \: \lambda   \\
      - sinh  \: \lambda  &  cosh  \:  \lambda
       \end{array}    \right)
\left(  \begin{array}{c}
         \tilde{\Phi}_{L}  \\   \tilde{\Phi}_{R}
       \end{array}   \right)
\end{equation}
with
\begin{equation}
tanh  \: 2\lambda = \frac{g_2}{2\pi v_F + g_4}
\end{equation}
By implementing this transformation to free boson fields
in the computation of
the fermion propagator we get
\begin{eqnarray}
\lefteqn{ \langle \Psi _R (x) \Psi _R ^{\dagger } (0)
      \rangle =  }     \nonumber      \\
& = &  \langle  e^{ i \Phi_R (x) }   e^{- i \Phi_R (0) }
      \rangle         \nonumber      \\
&  &     + \;  |c_1 |^2 e^{i 2 k_F x}
    \langle e^{i \left( \Phi_L (x) + \Phi_R (x) \right)}
    e^{i \Phi_R (x)} e^{-i \left( \Phi_L (0) + \Phi_R (0) \right)}
    e^{- i \Phi_R (0)}  \rangle    + \ldots  \nonumber   \\
& = &  \frac{d_1}{(2k_F x)^{1 + 2(sinh \: \lambda )^2}}
              \nonumber     \\
&  &  + \;  e^{i 2k_F x} \left[
\frac{d_2}{(2k_F x)^{3+2 (sinh \: \lambda )^2}} +
\frac{d_3}{(2k_F x)^{1 + 2 (cosh \: \lambda \: - \: sinh \: \lambda )^2
+  2 (sinh \: \lambda )^2}} \right]  + \ldots  \label{fin}
\end{eqnarray}
$d_1$, $d_2$ and $d_3$ are known constants whose explicit value is not
relevant for the purpose of the current discussion.
The important issue here is that the $2k_F$ oscillation,
which translates into a $3k_F$ oscillation of the electron propagator,
does not cancel out anymore. Only when
$sinh \: \lambda =0$ (free case) the cancellation takes place.
As we advanced previously,
the higher harmonic oscillations show up explicitly in the
interacting fermion propagator. In the
long distance limit and for $\lambda > 0$,
of the two terms
in the last line of (\ref{fin})
the second one
is actually the most relevant
as its exponent turns out to be smaller than that of the first.

To summarize,  we have found in this paper that the natural
way to understand the
emergence of higher harmonics in non-standard bosonization
formulas is the consideration of
a compact dispersion relation for the fermions in one dimension.
As a paradigm of this
situation we have taken an electron system on the lattice.
In our approach,
we have obtained the higher harmonics within a purely
kinematical framework. In this fashion we have
followed the standard bosonization procedure where the
boson representation is first
proposed for the free theory and, subsequently,
the interacting theory is solved without changing
the bosonization prescription.

As it is apparent from (\ref{ccr}) and (\ref{ccl}), our bosonization
formulas are quantitative in
the sense that we obtain explicit values for the
amplitudes associated to
higher harmonic terms.  The extension of the
present work to more complicated interacting fermion
systems and to fermions with spin is currently under study.

\end{document}